\documentclass[]{spie}
\usepackage[]{graphicx}  

\title{CZT in Space Based Hard X-ray Astronomy: Background Predictions for EXIST}

\author{A.~Garson~III\supit{a},
H. Krawczynski\supit{a}, 
G. Weidenspointner\supit{b}, 
E.I. Novikova\supit{c}, 
J. Grindlay\supit{d},\\
J. Hong\supit{d},
I.V. Jung\supit{a}
\skiplinehalf
{\small
\supit{a}Washington University in St. Louis, 1 Brookings Dr., CB 
    1105, St.\ Louis, Mo, 63130;
\supit{b}Centre d'Etude Spatiale des Rayonnements (CESR) 9 avenue du Colonel-Roche BP 4346, 31028, Toulouse, France;
\supit{c}Naval Research Laboratory, Washington DC;
\supit{d}Harvard-Smithsonian Center for Astrophysics, 60 Garden Street, Cambridge, MA 02138\\
}
}
\authorinfo{A. Garson III: agarson3@hbar.wustl.edu}
\begin{document}
\maketitle
\begin{abstract}

One of the key aspects of a detector material for space-borne hard X-ray and gamma-ray telescopes 
is the rate of prompt and delayed background events generated inside the material by charged 
and neutral particles striking the detector. These particles are Cosmic Rays, particles trapped in
Earth's magnetic field, and secondaries from Cosmic Ray interacting with the atmosphere and the spacecraft.
Here, we present a preliminary study of Cadmium Zinc Telluride (CZT) and its behaviour in space environments. 
We have used the simulation package MGGPOD to estimate the background of the CZT detectors 
in the proposed Energetic X-ray Imaging Survey Telescope ({\textit{EXIST}}) for possible
orbital parameters. The \textit{EXIST} mission will make use of $\sim$6~m$^2$ of $>$0.5~cm thick CZT detectors 
to record cosmic X-rays in the energy range from 10 keV to 600 keV. The detectors will be 
shielded by a fully or partly active shield. 
 For the specific detector and shielding geometry considered here and an orbit with a low (7$^\circ$) inclination, the background rate 
is dominated by diffuse extragalactic photons below $\sim$200 keV. Prompt and delayed hadronic backgrounds grow increasingly important above this energy, becoming the main contributors to the total background above $\sim$1 MeV. 
A fully active shield performs slightly better than a half active/half passive shield.

\end{abstract}


\section{Introduction}
With the launch of the Swift satellite in late 2004, CdZnTe (CZT) detectors entered the 
realm of space based X-ray and gamma-ray astronomy. Similar to CZT, CdTe has also been used as a detector material for instruments (IBIS and ISGRI) on the INTEGRAL mission prior to the operation of Swift. CZT offers many attributes that make 
it an excellent choice for a hard X-ray survey mission. Its wide band gap allows for 
operation without cryogenic cooling. Its high atomic number ($\bar{Z}\,=$ 50) results 
in high stopping power and a high photo-effect cross section. Furthermore, it offers 
good spatial and energy resolution from 10 keV to 600 keV.

The sensitivity of hard X-ray telescopes is limited by statistical fluctuations of the 
background rate, and thus the energy dependent background rate is an important
property of a detector material. One commonly distinguishes between prompt and delayed 
background events. Prompt background occurs when an incident particle immediately creates
a signal either by depositing energy or by causing a prompt nuclear reaction. 
All emission occurring within 1~$\mu$sec of an interaction is here considered 
to be prompt. Nuclear reactions can produce excited nuclei with long half-lives 
in the detector and surrounding materials. When these nuclei de-excite, their products can 
deposit energy in the detector producing a delayed signal. This process, 
referred to as activation, has a cumulative effect and the delayed background rates 
can increase during the lifetime of a mission.

There are several methods to suppress the background in space-borne telescopes. 
One is to passively shield the detector from incident particles, reducing the flux 
seen by the detector. While cost-effective and simple to design, there are limits 
to the impact of this technique. Above a certain energy (depending on shield material 
and geometry), particles will still be able to penetrate the shield (shield leakage) 
and interact with the detector. In addition, the shields will become activated 
from Cosmic Rays and neutrons, producing additional background events. 
In some cases, a better background suppression can be achieved with active shields 
that generate signals when hit. The veto-signals can be used to tag events if one or 
more particles deposit some of their energy in the shield. Furthermore, it is possible 
to tag ``internal background events'' generated inside the main detectors if high-energy 
photons leave the detector material and strike the shield. The tagged events can be 
excluded from the data analysis. Active shields require additional electronics and 
consume power. Some active shields add background themselves producing prompt or 
delayed gamma-ray emission. An optimal shield may use a combination of both, active and 
passive components.
\begin{figure*}
  \begin{center}
    \resizebox{!}{3in}{\includegraphics{{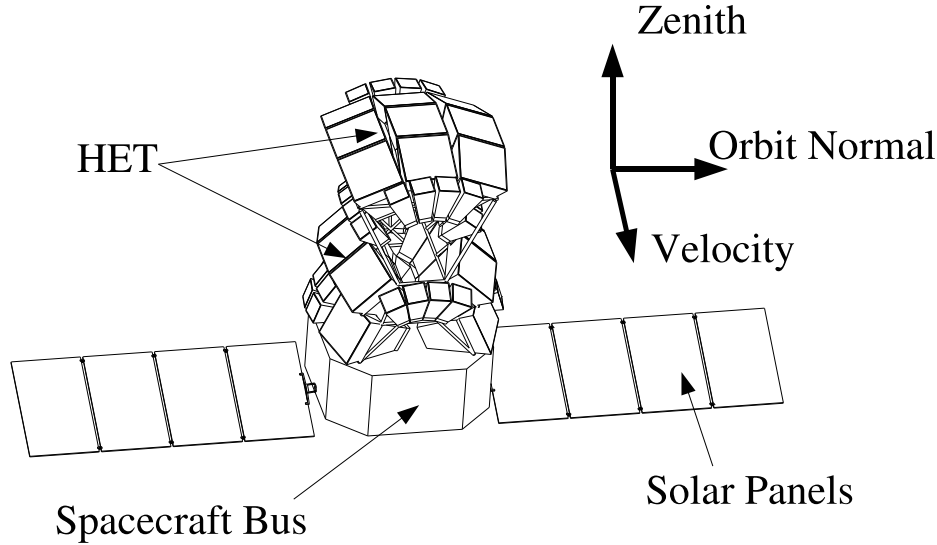}}}
    \caption{Concept design for the EXIST satellite that can monitor the entire X-ray sky every 95 min orbit. 
    The high energy telescopes (HETs) are sensitive from 10 keV - 600 keV. 
    The low energy telescopes (small boxes adjacent to the HETs, not labeled) detect photons with energy between 5 and 30 keV.}
  \end{center}
  \label{coffeemaker}
\end{figure*}

The \textit {Energetic X-ray Imaging Survey Telescope (EXIST)} is a proposed satellite-borne 
hard-X-ray telescope that can survey the entire 3-600 keV sky once every 95~min orbit 
(Grindlay et al. 2006). One design (Figure 1) combines a high-energy (HE) CZT detector 
with a low-energy (LE) Si detector, both being coded mask, wide field 
of view telescopes. The HE telescope will make use of $\sim$6~m$^2$ of $>$0.5~cm 
thick CZT detectors to detect 10-600 keV X-rays. The LE telescope complements the 
HE telescope by detecting 3-30 keV X-rays and improving the source localization accuracy 
from 1 arcmin to 10 arcsec. The scientific objective of {\it EXIST} is to survey the sky for 
the emission from obscured and non-obscured active galactic nuclei (AGNs), gamma-ray bursts (GRB), 
and galactic black hole systems. A unique feature of the mission with its large field of view 
would be the capability to discover rare transient phenomena, like the 
catastrophic event of a star plunging into a black hole. 
\textit{EXIST} was recommended as one of the three high energy missions in the 
2000 Decadal Survey Report, and is a candidate for the Black Hole Finder Probe (BHFP), 
one of three Einstein Probe missions recommended in NASA's Beyond Einstein Program.

We are carrying through a program to model the background in balloon and space-borne CdTe and 
CZT detectors. Our primary objective is to optimize the design of \textit {EXIST}. 
We plan to verify the modeling by simulating the background in the CZT and CdTe detectors 
of balloon and space borne-missions and by comparing the simulated with actually measured 
background rates and energy spectra. In this contribution, we describe our 
efforts to improve the simulation code and we show first preliminary results from 
simulating the background of the \textit {EXIST} CZT detectors. 
\section{Simulation Details}
\begin{figure*}
  \begin{center}
    \begin{minipage}[htb]{10.5cm}
    \hspace*{.05cm}
    \resizebox{!}{3in}{\includegraphics{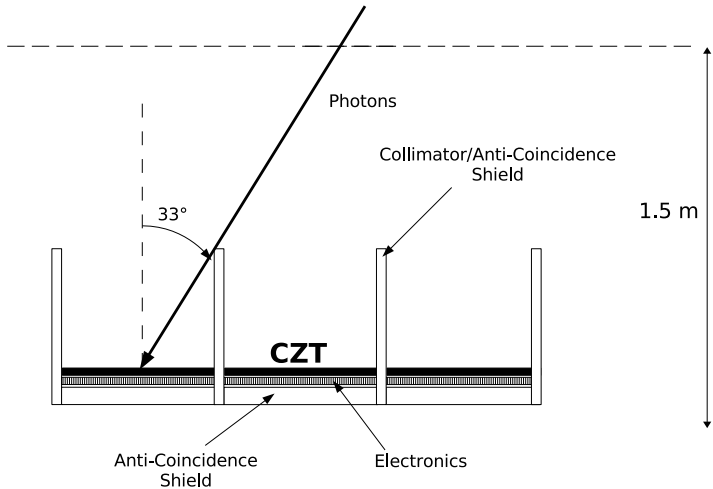}}
    \end{minipage}
    \begin{minipage}[htb]{6.5cm}
    \vspace*{-1cm}
    \caption{We have modeled this approximate geometry for {\it EXIST}, but the actual design may be different. The figure shows a side view of three high energy sub-telescopes with coded mask, CZT detectors, 
      and shield/collimator. The shadow pattern produced by the coded 
      mask is seen by the CZT detectors. 
      This pattern can be deconvoluted to reconstruct an X-ray image.
      The active shield/collimator defines the field-of-view for the 
      CZT detectors and reduces the instrumental background .}
    \label{box}
    \end{minipage}
  \end{center}
\end{figure*}

The \textit {EXIST} HE instrument is made of 19 sub-telescopes (see Figure 1). 
In rough approximation, we treat each sub-telescope as an rectangular box open on one side
and closed on 5 sides (see Figure 2).  In the following we refer to the open side 
of the collimator/shield assembly as ``top'' and the closed side as ``bottom''. 
Each sub-telescope uses a bottom shield and four side slabs for collimation and shielding. 
The bottom shield is always made of CsI(Na). We consider two different designs of the sides,
(i) fully active made of CsI(Na), and (ii) half active and half passive made of CsI(Na) and 
Pb-Sn-Cu. In the latter case, the active part surrounds the CZT detectors.
We use three different veto configurations (no veto, 100 keV threshold, and 1 MeV threshold).The bottom shield is active for all simulation runs using veto capabilities. 
While we assumed here a design based on right angles, the actual {\it EXIST} design 
joins shields at angles slightly larger than 90$^{\circ}$, resulting in a somewhat 
better photon detection efficiency. 

We modeled an approximate geometry for one {\it EXIST} HE sub-telescope, as shown in Figure2. Positioned directly above the bottom shield is a 1 cm 
thick region filled with readout electronics, above which lies a 0.5 cm thick CZT detector 
plane, both with an area of 56 $\times$ 56 cm$^{2}$. Located $\sim$1.5 m above the CZT is a 0.7 cm-thick 
tungsten coded mask. While the actual mask pixels will have a unit cell size of 2.5~ mm x 2.5~mm, 
we used a checkerboard pattern of tiles and gaps, each with an area of ~80 $\times$ ~80 cm$^2$, to reduce the simulation time. In addition, the coded mask used here provides similar detector response for alternating regions of the field of view (FoV). This is in contrast to the actual mask design which is optimized forboth uniform X-ray throughputs in the fully coded FoV (22$^{\circ}$ $\times$  22$^{\circ}$) and severe auto-collimation in the partially coded FoV.  For the diffuse cosmic X-ray background, whose event rates are strongly dependent on FoV, we approximate these effects with a slightly different mask model. The FoV up to  22$^{\circ}$ $\times$ 22$^{\circ}$ contains a checkerboard of 7~mm-thick tungsten mask and gaps with a unit cell size of ~14 cm $\times$  14 cm. The remainder of the FoV defined by the side shields/collimators (66$^{\circ}$ $\times$ 66$^{\circ}$ FWHM) is covered by continuous 3.5 mm-thick tungsten. This approximates the attenuation effect of the partly coded FoV and will be replaced in future simulations by an exact ray-tracing treatment for the radial mask holes. We use Al composite bars under the bottom shield to represent support structures 
for the sub-telescope. Finally, we introduce an Al-composite cube to represent 
the mass of the spacecraft. Table 1 gives the dimensions and properties 
of all components of the mass model. 

\begin{table}[t]
  
  \label{tb2}
  \centering
  \begin{tabular}{|c c c c|}
    \hline
    \textbf{Component} & \textbf{Material} & \textbf{Density} & \textbf{Dimensions} \\
     & & [g cm$^{-3}$] & l$\times$w$\times$t [cm$^{3}$] \\
    \hline
    \textbf{Detector} &  & &\\
    \hline
    CZT  & & 5.78 & 56 $\times$ 56 $\times$ 0.5 \\
    \hline
    \textbf{CsI(Na) Shields} & & & \\
    \hline
    Bottom  & CsI(Na) & 4.51 & 58 $\times$ 58 $\times$ 2\\
    Big Side  & CsI(Na) & 4.51 & 58 $\times$ 45 $\times$ 1 \\
    Small Side & CsI(Na) & 4.51 & 56 $\times$ 45 $\times$ 1 \\ 
    \hline
    \textbf{CsI(Na) +}&\textbf{Pb-Sn-Cu Shields} & & \\
    \hline
    Big Side Lower & CsI(Na) & 4.51 & 58 $\times$ 22.5 $\times$ 1 \\
    Big Side Upper & Pb-Sn-Cu & 10.5 & 58 $\times$ 22.5 $\times$ 0.5 \\
    Small Side Lower & CsI(Na) & 4.51 & 56 $\times$ 22.5 $\times$ 1 \\
    Small Side Upper & Pb-Sn-Cu & 10.5 & 56 $\times$ 22.5 $\times$ 0.5 \\
    \hline
    \textbf{Other} &  & &\\
    \hline
    Electronics & Si-composite & 1.8 & 56 $\times$ 56 $\times$ 1 \\
    Support Bars & Al-composite & 0.1667 & 58 $\times$ 1 $\times$ 1.5\\
    Spacecraft & Al-composite & 0.3675 & 124 $\times$ 124 $\times$ 124 \hspace*{0.5cm}\\
    Mask Tile (1 of 8) & Tungsten & 19.3 & 80 $\times$ 80 $\times$ 0.7 \\
    \hline
  \end{tabular}
  \caption{The densities (g/cm$^3$) and dimensions (length$\times$width$\times$thickness) are given for structures used in the sub-telescope mass model}
\vspace*{0.5cm}
\end{table}  

We use the MGGPOD software suite (Weidenspointner et al. 2005) to expose the mass model to the different radiation fields found in a low earth orbit. 
MGGPOD is comprised of five closely integrated packages: MGEANT simulates particle transport and electromagnetic interactions; GCALOR is responsible for handling hadron physics; PROMPT handles prompt emission from the de-excitation of nuclei due to spallation from incident protons and neutrons, inelastic neutron scattering, and neutron capture; the ORIHET package calculates the build up and decay of activity based on nuclide production rates; DECAY then simulates the radioactive decays.
MGGPOD results match well with the experimental data from the Transient Gamma-Ray Spectrometer (TGRS) experiment \cite{Weid}. 

An important part of simulating the delayed background is the consideration of hadronic interactions. The GCALOR code consists of 4 packages whose use depends on incident particle and energy. Table 2 summarizes these packages and their relevant particles and energies. Of special consideration is the MICAP energy regime, as all physics of neutron interactions must be provided via cross-section files.  
\begin{table}
  \centering
    \begin{tabular}{|l| l l l l l l l l |l|} 
      \hline
	Element& \multicolumn{8}{|c|}{Isotopes} & Data Base \\ \hline 
	Na & $^{23}$Na & & & & & & & & JENDL-3.3\\ 
Al &$^{27}$Al & & & & & & & & ENDF/B-VI.8\\ 
Ti &$^{46}$Ti &$^{47}$Ti &$^{48}$Ti &$^{49}$Ti & $^{50}$Ti& & & & JENDL-3.3\\ 
Cu &$^{63}$Cu & $^{65}$Cu& & & & & & & JENDL-3.3\\ 
Cd &$^{106}$Cd &$^{108}$Cd &$^{110}$Cd &$^{111}$Cd & $^{112}$Cd&$^{113}$Cd &$^{114}$Cd & $^{116}$Cd &  ENDF/B-VI.8\\ 
Te &$^{120}$Te &$^{122}$Te &$^{123}$Te &$^{124}$Te & $^{125}$Te&$^{126}$Te &$^{128}$Te & $^{130}$Te & JENDL-3.3\\ 
W &$^{182}$W &$^{183}$W &$^{184}$W &$^{186}$W & & & & & JENDL-3.3\\ 
Pb &$^{204}$Pb & & & & & & & & JENDL-3.3\\
Pb &$^{206}$Pb &$^{207}$Pb &$^{208}$Pb & & & & & & ENDF/B-VI.8\\ 
Bi & $^{209}$Bi & & & & & & & & ENDF/B-VI.8\\ 
I & $^{127}$I & & & & & & & & JENDL-3.3\\ 
Cs & $^{133}$Cs & & & & & & & & JENDL-3.3\\ 
Zn & $^{64}$Zn&  $^{66}$Zn &$^{67}$Zn &$^{68}$Zn &$^{64}$Zn &  $^{70}$Zn &(Derived) & & JEFF-3.1/A Brond-2.2\\ \hline
\end{tabular}
\caption{Elements and their isotopes for which neutron cross section reformatted and derived (Zn) from the officially available data bases.}
\label{tab:elements}
\end{table} 

Many isotopes of interest had been missing from the original GCALOR/MICAP neutron cross section data banks. New cross sections data files were developed by processing data from JENDL-3.3 and ENDF/B-VI.8 data sets to make them compatible with the MICAP format. In addition, element Zn is missing from the JENDL-3.3 and ENDF/B-VI compilations. Additional neutron cross section files were assembled based on data from the JEFF-3.1A, BROND-2.2, and EXFOR data bases. 
Table 2 lists the isotopes for which cross section files were compiled.
\begin{table}
\centering
\begin{tabular}{|l| l  |} 
\hline
MGEANT& Particle transport, electromagnetic interactions \\ \hline 
GCALOR & CALOR package for hadron physics \\ 
&1. NMTC (Nucleon Meson Transport Code)\\
&$\:\:\:\:\:\:$nucleons 1 MeV to 3.5 GeV, charged pions 1 MeV to 2.5 GeV\\
& 2. Scaling Model\\
&$\:\:\:\:\:\:$nucleons and charged pions 3 GeV to 10 GeV\\
&3. FLUKA\\
&$\:\:\:\:\:\:$nucleons and charged pions above 10 GeV and for all energies for particle types\\
&$\:\:\:\:\:\:$not implemented in other portions\\
&4. MICAP\\
&$\:\:\:\:\:\:$neutrons $1\cdot 10^{-5}$eV to 20 MeV\\ \hline
PROMPT&Simulates prompt emission from de-excitation of nuclei due to spallation from \\
& incident protons and neutrons, inelastic scattering and neutron capture\\ \hline
ORIHET & Calculates build up and decay of activity based on nuclide production rates\\ \hline
DECAY&Simulates radioactive decays\\ \hline
\end{tabular}
\caption{This table describes the five main software components incorporated into MGGPOD. The four software packages used by GCALOR are also given. For each GCALOR package, the relevant particles, interactions, and energy regimes are given.}
\label{tab:MGGPOD}
\end{table} 

\section{Background Input Spectra}
We consider six background components relevant to a 550 km orbit at 7$^\circ$ inclination, other orbital inclinations will be considered later. This orbit is realistic for a spacecraft weight of 8500 kg and a launch vehicle of the class Delta IV(4050H). Each component is simulated separately, with the final results being combined to form a total background. The particles are generated with a random position and trajectory on the surface of a cylinder with radius 1500 cm and height 2000 cm, centered about the sub-telescope. Prompt energy deposits in the CZT were recorded and filled into histograms, weighted for a 1 sec interval. At the same time, the production of excited nuclei was also recorded. Any prompt de-excitations were included in the prompt deposited energy histograms. The resulting nuclei production rates were then calculated and evaluated for various orbital histories. The nuclei were allowed to de-excite and decay for 30 sec, while the resulting delayed energy depositions in the CZT were recorded.  

The simulated background components include diffuse extragalactic X-rays, cosmic ray protons and electrons, 
trapped protons, and atmospheric neutrons. We use the diffuse extragalactic X-ray background as measured by the HEAO-1 satellite (Gruber et al. 1999). A flux of atmospheric neutrons (ANs) is produced from cosmic ray interactions. We use one of two input spectra, depending on neutron energy: COMPTEL observations from Morris (95,98) above 10 MeV, and theoretical calculations from Lingenfelter (1963) below 10 MeV. ESA's SPace ENVironment Information System (SPENVIS)\footnote{http://www.spenvis.oma.be} determined the trapped proton peak fluxes incorporated in the simulations and CREME96 \footnote{https://creme96.nrl.navy.mil/} computed the time averaged cosmic ray proton (CRP) flux. We assume the instruments are turned off during 
passages through the South Atlantic Anomaly (SAA), so only delayed events are considered for trapped protons. 
Using the orbital parameters from above, the results from SPENVIS show that SAA passages occur 
typically 3 - 4 times per day, with an average duration of 10 minutes. 
For our simulations, we assumed SAA passages occur every 6 hours for 10 minutes. 
Finally, the cosmic electron flux from Gehrels (1985) was included with a low-energy cutoff at 2.5 GeV.

\section{Results}
The average delayed background rates are expected to increase during the first weeks and months of the mission. 
After roughly half a year, the background rates reach a rather stable mean level with short-time variations
owing to SAA passages. Figure 3 shows the total (prompt+delayed) background rate in the CZT 
detectors one year into the mission. We assumed a fully active shield and show the total rate 
before and after applying a veto with an energy threshold of 100 keV deposited in any one shield. 
The vetoes reduce the background rate above 200 keV by a factor of 2.
The figure also shows the relative contribution from different background components (before veto).
The dominant sources are diffuse photons and cosmic ray protons. 
In our model, diffuse photons are most important up to $\sim$200 keV. 
 At higher energies, hadronic prompt and delayed events grow increasingly important. 
Above $\sim$2 MeV, the CRP prompt events become dominant. 
The cosmic electron event rate is negligible at all energies of interest. The total un-vetoed event rate between 10 and 600 keV is $\sim$0.45~cm$^{-2}$ s$^{-1}$. 

Figure 4 shows that the delayed background originates mainly in the 
CZT detectors and in the shield (fully active shield, after shield-veto, 
veto energy threshold of 100 keV). The mask, electronics, and support 
structures contribute little to the delayed signal. The masks are 
too far away from the CZT detectors. The effects from activation of the
support structures and electronics is limited due to shielding by the CsI(Na). 

\begin{figure*}
  \begin{center}
    \begin{minipage}[htb]{9.5cm}
    \hspace*{-0.0 cm}
    \resizebox{!}{2.45in}{\includegraphics{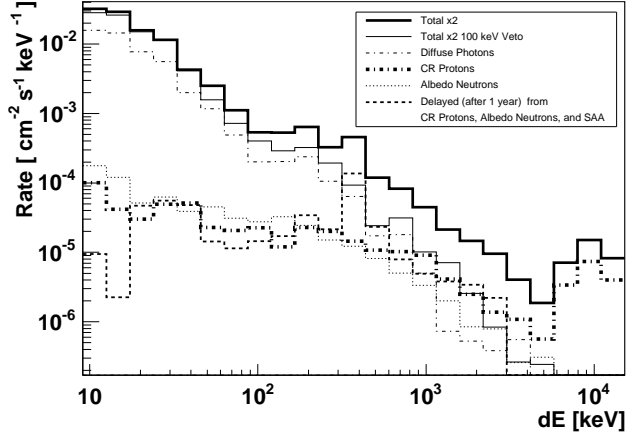}}
    \end{minipage}
    \begin{minipage}[htb]{6.5cm}
    \vspace*{-.1cm}
    \caption{Total Background (prompt+delayed) before and after applying a shield-veto whenever an event deposits $>$100 keV 
in any one of the shields (fully active configuration). The relative contribution from 
different environmental components are also shown (before veto). 
We assumed an in-orbit time of 1 year with activation by primary CR protons, albedo neutrons, 
SAA particles, and secondary particles produced by interactions of the primary particles 
with the satellite. The figure assumes that 5 min passed since the last SAA passage.
}
    \label{bg}
    \end{minipage}
  \end{center}
\end{figure*}
\begin{figure*}
  \begin{center}
    \begin{minipage}[htb]{9.5cm}
    \hspace*{-0.25cm}
    \resizebox{!}{2.45in}{\includegraphics{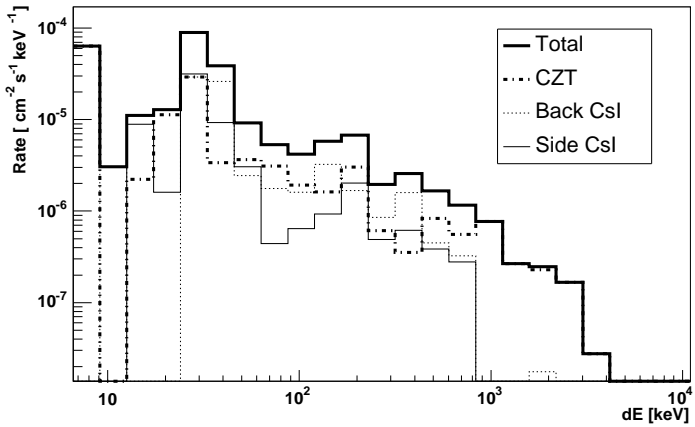}}
    \end{minipage}
    \begin{minipage}[htb]{6.5cm}
    \vspace*{-1cm}
    \caption{Structural contributions to the delayed background due to one year exposure to CR protons and albedo 
neutrons (after shield-veto). These results are for a fully active shield with a 100 keV veto energy threshold. 
The CZT detectors and CsI(Na) shields are significant sources of delayed background events.
The contribution from the coded mask, electronics, and support structures are negligible and 
are not included in this figure.  }
    \label{st}
    \end{minipage}
  \end{center}
\end{figure*}

We estimated the delayed signal for different durations in orbit. 
Figure 5 shows the background rates for one hour, one day, six months, one year, 
and two years of exposure to CRPs, ANs, and their secondaries
(fully active shields, 100 keV veto energy threshold). 
One can see the large increase in rate between one month and six months, 
however, the difference between one year and two years is not as dramatic, 
as nuclei production and decay rates tend towards equilibrium. This is expected as there are only a few isotopes with lifetimes longer than one year.
The same figure shows the delayed event rate from activation by 
SAA particles and their secondaries 5 min after the last SAA passage. 
All three activation components (CRPs, ANs, SAA) are important
and contribute to the total rate of delayed emission.

 Figure 6 compares the ``fast'' and ``slow'' components of the SAA delayed background. Here, fast (FC) refers to the events that are products of secondary nuclei with short ($<$1 Day) half-lives, while events from excited nuclei with longer half-lives are deemed slow (SC). For one SAA passage, the event rates are shown for various cool down times after the passage (CsI(Na) shields with no veto). The cool down times range from immediately after exiting the SAA to 24 hours. The long component represents 1 year of SAA exposures. These results show that one year into the mission, the long-lived component produces a similar background rate as the short-lived component.

\begin{figure*}
  \begin{center}
    \begin{minipage}[htb]{9.5cm}
    \hspace*{-0.25cm}
    \resizebox{!}{2.5in}{\includegraphics{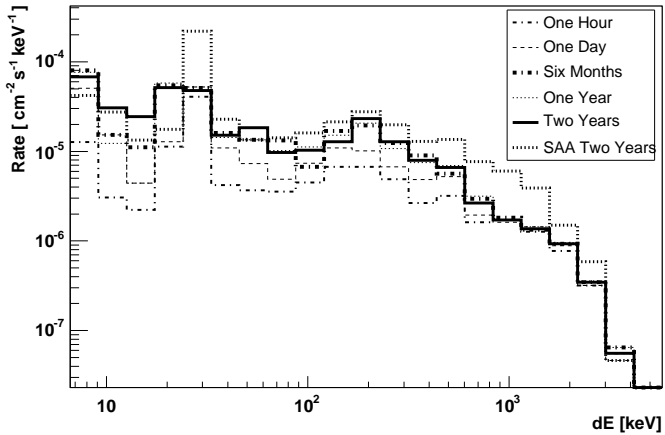}}
    \end{minipage}
    \begin{minipage}[htb]{6.5cm}
    \vspace*{-1.5cm}
    \caption{The delayed background resulting from cosmic ray proton and albedo neutron interactions 
are shown for different durations in orbit. The background rates increase for increasing durations in 
orbit, but tend to level off by six months. Also shown is the delayed background due to two 
years of SAA passages 5 min after the last SAA passage.
}
    \end{minipage}
    \label{Dt}
  \end{center}
\end{figure*}

 \begin{figure*}
  \begin{center}
    \begin{minipage}[htb]{9.5cm}
    \hspace*{-0.25cm}
    \resizebox{!}{2.5in}{\includegraphics{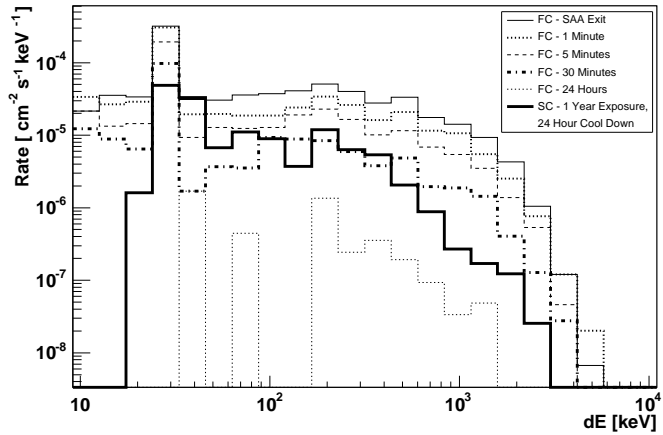}}
    \end{minipage}
    \begin{minipage}[htb]{6.5cm}
    \vspace*{-0.5cm}
    \caption{This figure shows delayed background rates produced by a single SAA passage for different ``cool down times'': just exiting the SAA (solid thin line), 1 min (dotted-thick line), 5 min (dashed line), 30 min (dot-dash line), and 24 hours (dotted-thin line) after exiting the SAA. We compare these background rates with the long-lived activation from the SAA passages accumulating over one year in orbit(solid-thick line),shown here 24 hours after the last passage. These results assume only passive CsI(Na) shielding. }
    \label{VP}
    \end{minipage}
  \end{center}
\end{figure*}

\begin{figure*}
  \begin{center}
    \begin{minipage}[htb]{9.5cm}
    \hspace*{-0.25cm}
    \resizebox{!}{2.5in}{\includegraphics{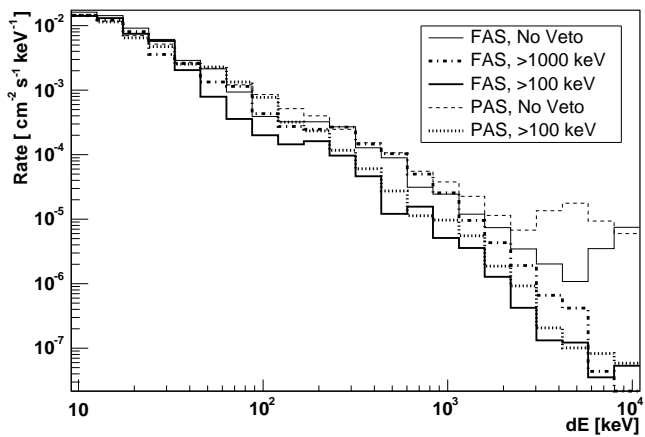}}
    \end{minipage}
    \begin{minipage}[htb]{6.5cm}
    \vspace*{-0.5cm}
    \caption{The total prompt and delayed background rate is shown for five different shield/veto-energy-threshold configurations.
FAS stands for ``fully active shield'' and PAS stands for ``partly active shield''. 
The delayed components include contributions from CR protons and albedo neutrons after one year 
in orbit, as well as SAA effects from the six most recent passages (30 hours) in orbit, 5 min after the last SAA passage. }
    \label{VP}
    \end{minipage}
  \end{center}
\end{figure*}

Figure 7 displays the combined prompt and delayed (after one year) background rates using four different shielding scenarios;  fully active (CsI(Na)) shield with no veto, veto with 100 keV and 1000 keV energy threshold, partly active (CsI(Na)+Pb-Sn-Cu) shield with no veto and veto with 100 keV energy threshold.   
Delayed CZT background rates resulting from  CRP, AN, and delayed SAA (last six passages, 5 minutes 
after last passage) events are given. As expected, using a ``low'' 100 keV shield threshold minimizes 
the background rate for both material configurations. The fully active shield (FAS) provides a 
lower background rate over the energy range of interest. For some energies, the detected rate using shield veto is higher than without. This is an effect from simulating with different incident photon sets.    
\section{Conclusions}
The preliminary results from our simulations indicate that active shielding is an effective 
means of reducing the background for the \textit{EXIST} mission. Fully active shields 
appear to perform better than partly-active/partly-passive and fully passive shields.
For our detector, mask, and shielding geometry, and assuming a low 7$^{\circ}$ inclination orbit, the \textit{EXIST} background seems to be 
dominated by aperture flux (photons entering the detector assembly through
the optical path), and shield leakage (photons passing through the shield without causing a shield-veto) below 200 keV. Above $\sim$1 MeV, the prompt and delayed signals from hadronic background components are the main contributors to the total background rates.
Future work will build on these first results, incorporating more detailed 
mass models and environmental conditions. Other detector and shielding configurations will be evaluated. Another possible design uses more collimation and canted side shields. We plan to investigate further the effects from different ratios of active/passive shielding, as well as evaluating the pay-off of 
using thicker shields to reduce the shield leakage. Furthermore, we plan to use the MGGPOD code to simulate the background in 
the CZT/CdTe detectors of the existing missions Swift and INTEGRAL 
and of past balloon-borne experiments. Comparison of simulated with 
experimentally measured background rates are instrumental for checking 
the simulation tools and the input spectra.

\acknowledgements{
  This work is supported under NASA grant NNH05ZDA001N and the McDonnell Center for the 
  Space Sciences at Washington University in St. Louis.}

\end{document}